\documentclass[%
 aip,
 amsmath,amssymb,
reprint,%
]{revtex4-1}

\usepackage{graphicx}
\usepackage{dcolumn}
\usepackage{bm}

\usepackage[utf8]{inputenc}
\usepackage[T1]{fontenc}
\usepackage{mathptmx}
\usepackage{physics}

\usepackage{comment} 
\usepackage{xcolor} 

\includecomment{sjsComment}
\newcommand{\sjs}[1]{}
\begin{sjsComment}
\renewcommand{\sjs}[1]{\textcolor{red}{[sjs: #1]}}
\end{sjsComment}

\begin{document}

\preprint{AIP/123-QED}

\title[Adaptive-Cooling Simulated Annealing]{Simulated Annealing with Adaptive Cooling Rates}

\author{Mariia Karabin}
\email{mariya.karabin8@gmail.com.}
\author{Steven J. Stuart}%
 \email{ss@clemson.edu.}
\affiliation{Department of Chemistry, Clemson University, Clemson, SC 29634 USA
}

\date{\today}

\begin{abstract}
As one of the most robust global optimization methods, simulated annealing has received considerable attention, with many variations that attempt to improve the cooling schedule.
This paper introduces a variant of simulated annealing that is useful for optimizing atomistic structures, and makes use of the statistical mechanical properties of the system, determined on the fly during optimization, to adaptively control the cooling rate.
The adaptive cooling approach is demonstrated to be more computationally efficient than classical simulated annealing, when applied to Lennard-Jones clusters. 
This increase in efficiency is approximately a factor of two for clusters with 25--40 atoms, and improves with the size of the system. 
\end{abstract}

\maketitle

\section{\label{sec:intro}Introduction}

Simulated annealing (SA) is one of the most robust optimization methods, and has been widely used for global optimization and energy minimization problems for decades\cite{led08}. 
The method makes use of Boltzmann sampling of an energy landscape at a temperature which is gradually reduced\cite{Leach01,kir83}.
At elevated temperatures, the system has enough energy to be able to cross energy barriers and find the basins that contain the important energy minima, and at low temperatures (with suitably slow cooling) it converges to a neighborhood around the global minimum. 

The simulated annealing method is widely used, due to its simplicity and ease of implementation, but suffers from being very time consuming\cite{Fren12}.
It is guaranteed to converge to the global minimum only in the limit of infinitely slow cooling rates, so a very gradual decrease in temperature gives the best results, but is also computationally costly.
If the temperature is decreased too rapidly, the system is  likely to become trapped in a local minimum\cite{har16}.

For this reason, considerable effort has been devoted to improving the efficiency of the SA algorithm,\cite{sid16} based on improvements to the cooling schedule, learning mechanism, and neighborhood selection.
For example, Monte Carlo sampling has been performed with wide-tailed distributions for sampling states,\cite{szu87}
and hybrid methods have been developed that extend the Boltzmann or Monte Carlo sampling used in traditional SA to include sampling based on a genetic algorithm\cite{lee15}, differential evolution\cite{vin15}, particle swarm optimization\cite{sha15}, and the harmony search algorithm\cite{gee01}.

Independently from the method used to sample states in configuration space, the cooling schedule used to control the annealing temperature can also have an effect on the efficiency of the optimization.
The goal is to design the fastest cooling schedule that still results in an acceptably high probability of convergence to the global minimum.
Much effort has been devoted to studying whether the optimal cooling schedule is linear in time, or exponential\cite{ing89},  proportional\cite{kir83}, nonlinear\cite{lun86},
inversely linear\cite{szu87},
logarithmic\cite{gem84},geometrical\cite{nou98}, or adaptive\cite{fod12,ing93}.
Adaptive cooling schedules have also been developed that use information about the system such as the variance of the energy or the temperature to adjust the cooling\cite{sid16,van92}.
Unfortunately, the answer to this question depends on the system.
Systems with a folding funnel-style energy landscape can often tolerate more rapid cooling, while systems with a rugged energy landscape and a broad distribution of energy barriers may need a more conservative cooling schedule.

Slow cooling is most important at temperatures where the thermal energy $kT$ allows only infrequent crossing of the critical barriers allowing escape from local minima.
An ideal cooling schedule would cool quickly during portions of the optimization when the system is unlikely to become kinetically trapped in a local minimum, but much more slowly at key temperatures where kinetic traps are more accessible.
Unfortunately, this idealized approach requires detailed knowledge of the features of the function being optimized --- features which are generally not known if the function is being optimized.

What is needed is a method that can adaptively determine the instantaneous cooling rate, based on the energetic properties of the system being simulated, but without making use of any a priori knowledge of the features of the energy landscape.
The purpose of this paper is to describe such a method, using on-the-fly statistical mechanical property evaluation to adaptively adjust the cooling rate in real time during the optimization.
Ideally, these properties would include a weighted distribution of barrier heights accessible at the current thermal energy, but this requires non-local knowledge of the potential energy surface.
As a proxy for this information, we use the heat capacity of the system, which is largest at temperatures where a range of new energy basins is just becoming populated.

Thus, when the heat capacity is large, a slow cooling rate is used to ensure that the system spends proportionally more time exploring configuration space and crossing energy barriers, reducing the probability that the system becomes kinetically trapped in a local minimum.
When the heat capacity is small, on the other hand, the system can be cooled at a much faster rate.
This decreases the computational time, with comparatively lower risk of becoming trapped in an undesirable minimum.

For example, consider the case of a phase change, such as the liquid-solid transition that originally inspired the simulated annealing method.
To anneal a liquid into the global-minimum solid configuration, rather than one of the many local-minimum glassy configurations, it would suffice to cool very rapidly to just above the melting temperature, then very slowly across the phase transition, and then very rapidly thereafter.
The heat capacity becomes very large near the phase transition temperature, so this can be used as a signature that slow cooling is needed, even when the phase transition temperature is unknown ahead of time.
Note, also, that the cooling rate varies quite non-monotonically, alternating between slow and fast cooling rates.
Even in the absence of a bona fide phase transition, the heat capacity will be larger at temperatures where more states are becoming thermally accessible, and these are the temperatures where a SA algorithm should be cooling most slowly.

\section{Methods}

In the most basic implementation of SA,\cite{kir83} the temperature is lowered from some initial temperature, $T_i$, to a final temperature, $T_f$, using an exponential cooling schedule,
\begin{eqnarray}
T(t) &=& T_i e^{-kt},
\end{eqnarray}
with  a constant cooling rate $k$.
The varying temperature is used to perform Boltzmann sampling of the states of the system.
Both the success and the efficiency of the optimization depend strongly on this cooling rate; when $k$ is too large, the system will quench into a non-global minimum with an unacceptably large probability, but when $k$ is too small, the optimization will be unacceptably slow to complete.
Unfortunately, $k$ must be chosen before the optimization begins, and usually before much is known about the distribution of local minima on the energy landscape.
Consequently, $k$ is usually treated as a purely empirical parameter; it is chosen to be as small as can be computationally afforded, in the hopes that this will find the global minimum.

We propose a modification of this classical SA algorithm, in which the cooling rate varies with the progress of the optimization, $k(T)$.
The annealing schedule becomes a complicated function of the history of the past cooling rates,
\begin{eqnarray}
T(t) &=& T_i e^{-\int k(T(t)) \, \mathrm{d}t},
\end{eqnarray}
but the actual cooling can be implemented quite easily using finite-difference decrements in the  temperature using the instantaneous cooling rate,
\begin{eqnarray}
T(t + \Delta t) &=& T(t) - k(T) \Delta t. \label{eq:cool}
\end{eqnarray}

In principle, the cooling schedule $k(T)$ could be an arbitrarily complicated function; the optimal cooling schedule would be different for every system, and difficult to obtain.
But the main intent of the variable cooling rate is to have the optimization proceed slowly only as the system cools across important transition temperatures, while cooling more rapidly away from these temperatures.
Consequently, we propose a dual cooling rate approach, in which the cooling occurs at a fixed, slow rate, $k_s$, when the instantaneous heat capacity of the system is above some cutoff, $C_V^*$, and a different, faster rate, $k_f$, when the heat capacity is below the cutoff:
\begin{eqnarray}
k(T) &=& \left\{ \begin{array}{rl}
k_s, & C_V(T) \ge C_V^* \\
k_f, & C_V(T) < C_V^*
\end{array} \right. \label{eq:k}
\end{eqnarray}

To evaluate the heat capacity with no a priori information about the system, we make use of the fluctuation formula,
\begin{eqnarray}
 C_V &=& \frac{ \left<E^2\right> - \left<E\right>^2}{k_B T^2}, \label{eq:CV}
\end{eqnarray}
in the canonical ensemble, where  $\left< E \right>$ and $\left< E^2 \right>$ are evaluated at a fixed $T$.

Thus, a full implementation of the dual-cooling rate simulated annealing (DRSA) algorithm involves the following steps:
\begin{enumerate}
    \item Begin at temperature $T_i$.
    \item Equilibrate the system by sampling $N_{\mathrm{eq}}$ steps in the canonical ensemble at the current temperature. \label{step:eq}
    \item Sample $N_{\mathrm{prod}}$ steps in the canonical ensemble at the current temperature, and use these to evaluate $C_V$ (Eq. \ref{eq:CV}) and the associated cooling rate (Eq. \ref{eq:k}).
    \item Cool the system for $N_{\mathrm{cool}}$ steps using the current cooling rate (Eq. \ref{eq:cool}).
    \item End the simulation, if the temperature has fallen to $T_f$; otherwise return to step \ref{step:eq}.
\end{enumerate}
We will refer to this modification of the SA algorithm as adaptive-cooling simulated annealing, or ACSA.
Note that the method is characterized by eight different parameters: $T_i$, $T_f$, $k_s$, $k_f$, $C_V^*$, $N_{\mathrm{eq}}$, $N_{\mathrm{prod}}$, and $N_{\mathrm{cool}}$.
(The classical SA method requires only $T_i$, $T_f$, and $k$.)

The requirement to evaluate the heat capacity at a fixed temperature builds some inefficiency into the optimization algorithm.
Only some of the total sampling steps are used to cool the system; this productive fraction of the simulation is
\begin{eqnarray}
f &=& \frac{N_{\mathrm{cool}}}{N_{\mathrm{eq}} + N_{\mathrm{prod}} + N_{\mathrm{cool}}}.
\end{eqnarray}
The remaining portion of the steps ($1-f$) represent the computational overhead required to evaluate the heat capacity using Eq.~(\ref{eq:CV}).
The expectation is that the faster cooling rate $k_f$ applied during some portions of the optimization will more than compensate for this overhead.

To quantify the performance of the ACSA algorithm, and compare it to classical SA, we measure both the computational cost of the  optimization as well as its accuracy.
The computational cost, $N$, is the total number of sampling steps taken during the optimization.
For classical SA, this can be determined directly from the cooling rate, along with the starting and ending temperatures: $N = \frac{1}{k \Delta t} \ln \frac{T_i}{T_f}$.
For ACSA, it depends on the heat capacities calculated from the sampled states, and may be different for different trials.
In practice, we measure the mean cost, $\left< N \right>$, across many optimizations.

The accuracy of the optimization is determined by the probability, $p$, that an optimization finishes in the global-minimum energy basin.
For this reason, all of the optimizations here are performed on LJ clusters for which the global minimum energy configuration is already known\cite{wal97}.
A configuration of an LJ cluster is defined to be in the global minimum basin if each of the interatomic distances $\left\{ r_{ij} \right\}$ is identical to those in the global minimum energy configuration, to within a tolerance of $r_{\mathrm{min}} = \pm 0.24$ (reduced units).
To estimate the accuracy of an optimization algorithm, it is run for $M$ independent trials, using different pseudo-random number seeds.
If $M_{\mathrm{succ}}$ of the trials terminated in the global minimum energy basin, and $M_{\mathrm{fail}} = M - M_{\mathrm{succ}}$ do not, then the success probability is estimated as $\left< p \right> = M_{\mathrm{succ}} / M$.
Because this is a binomial process, the standard error of the measured mean accuracy is
\begin{eqnarray}
\sigma_p = \sqrt{\frac{1}{M}p \left( 1 - p \right)}.
\end{eqnarray}

These two metrics, $p$ and $N$, characterize the performance of an optimization method.
Ideally, one would prefer a method with high $p$ and low $N$.
In practice, however, the accuracy is improved by reducing the cooling rate(s), which comes with increased computational cost.
The important question, then, is how much computational effort is worth investing to improve the accuracy of the optimization method.
This question has a natural answer when we observe that an inaccurate optimization method can be used to find the global optimum quite accurately when it is repeated in multiple independent trials.

For example, suppose an algorithm predicts a global minimum with some probability $p$.
Even if $p$ is not close to 1, so that there is a substantial probability that a single trial will fail to find the global minimum, there may still be a large probability that {\em most} of the trials will succeed.
If we run this algorithm 3 times, independently, the probability that the method terminates in the global minimum more than half of the time (i.e. in 2 or 3 trials) is $p^3 + 3p^2(1-p) = p^2(3-2p)$.
More generally, for $n$ independent trials, the probability that at least half of the trials succeed is
\begin{eqnarray}
p_n &=& \sum_{k > n/2} {n \choose k} p^k \left( 1 - p \right)^{n-k}. \label{eq:majority}
\end{eqnarray}
As long as $p > \frac{1}{2}$, the accuracy $p_n$ increases monotonically with $n$.
This technique is often used in practice to obtain useful results from an imperfect optimization method; if repeated optimizations identify the same final state, it can be declared the global minimum with more confidence than can the result of a single optimization.

Note that we intentionally require that the {\em majority} result match the global minimum, rather than the {\em lowest} energy obtained from multiple trials.
This is because optimization algorithms are often applied to many-dimensional systems, which suffer from the curse of dimensionality.
The number of states that are thermally accessible within an energy of $kT_f$ above a local minimum scales as $T_f^{d/2}$ for a $d$-dimensional system.
When a many-dimensional optimization is halted at a temperature $T_f$ low enough to have localized to a particular energy basin, but above 0~K, it is extremely unlikely to be found at the true minimum.
If there are local minima with energies within a few $kT_f$ of the global minimum, then there is no guarantee that the lowest final energy is in the basin with the lowest minimum energy.
If such distractor minima are not a problem, and it suffices to find the global energy basin a single time, then the $n$-trial success criterion in Eq.~(\ref{eq:majority}) can be replaced with
\begin{eqnarray}
p_n &=& \sum_{k=1}^n {n \choose k} p^k \left( 1 - p \right)^{n-k} \\
&=& 1 - (1 - p)^n.
\end{eqnarray}
This metric increases much more rapidly with $n$ than does the majority-based $p_n$.

Note also that requiring a majority of the optimizations to find the global-minimum basin is overly conservative.
If there are more than two local minima, and if the final states can be accurately clustered into distinct basins, then all that is needed is that a {\em plurality} of the final states be in the global energy basin.
The accuracy metric in Eq.~(\ref{eq:majority}) could be replaced with its multinomial equivalent,
\begin{eqnarray}
p_n &=& \sum_{k_0 > k_j, \forall j} {n \choose k_0 k_1 \cdots} \prod_j p_{(j)}^{k_j} \label{eq:multinomial}
\end{eqnarray}
However, this requires that the probabilities $p_{(j)}$ be known for each of the distractor minima, and Eq.~(\ref{eq:multinomial}) does not yield as easily to a continuous approximation, as discussed below, so we have not pursued this approach further.

For a large number of trials, the cumulative distribution function in equation~\ref{eq:majority} is tedious to evaluate by direct summation.
It is equivalent, and more convenient for large $n$, to evaluate $p_n$ using the incomplete beta function.
Assuming odd $n$, 
\begin{eqnarray}
p_n &=& I_p\left(\frac{n+1}{2},\frac{n+1}{2}\right) \nonumber \\
&=& \frac{\int_0^p t^{\frac{n-1}{2}} \left( 1-t \right)^{\frac{n-1}{2}} \dd{t}}{\int_0^1 t^{\frac{n-1}{2}} \left( 1-t \right)^{\frac{n-1}{2}} \dd{t}} \label{eq:majorityBeta}
\end{eqnarray}

Equation~\ref{eq:majority} or \ref{eq:majorityBeta} describes the probability that an optimization, with single-trial success rate $p$, will arrive at the correct consensus result after $n$ trials.
An optimization approach with poor single-trial accuracy can be improved with repetition, and $p_n$ describes exactly how fast the accuracy improves with repeated trials.
This gives us the basis for defining a single scoring metric that combines accuracy and computational cost.

Consider a SA optimization method with accuracy $p$ and average computational effort $N$.
To improve the accuracy, either the cooling rate can be reduced, or the optimizations can be repeated for multiple trials.
Using $n$ trials will improve the accuracy to $p_n$, at a cost of increasing the computational effort to $nN$.
A modification to the optimization algorithm that delivers a higher accuracy of $p' = p_n$ in a single trial, should do so at a computational effort of less than $nN$, or else we would prefer repeated trials of the original optimization.
In other words, an optimization method characterized by accuracy and effort $(p,N)$ is equivalent to one characterized by $(p_n,nN)$.

To quantify this relationship, let $\nu(p,p')$ be the number of times that a method with accuracy $p$ would need to be repeated in order to achieve accuracy $p'$.
That is, $p_{\nu(p,p')} = p'$.

This definition assumes $p \le p'$.
In cases where $p > p'$, the definition is extended so that 
\begin{eqnarray}
\nu(a,b) &=& \frac{1}{\nu(b,a)}.
\end{eqnarray}
In this way, $\nu$ can be interpreted as a real-valued measure of the relative cost of achieving two different accuracies (through repeated trials).
(Note that although the number of repeated trials in Eq.~\ref{eq:majority} must be an integer, $n$ has been extended to real values in Eq.~\ref{eq:majorityBeta}.)
For example, $\nu(0.90,0.972) = 3$ because three repeated trials with accuracy 0.9 achieve a consensus accuracy of 0.972.

The efficiency of two methods with different accuracy and cost can be ranked by comparing the cost for each to obtain some benchmark accuracy.
The computational effort required to achieve a target accuracy $\alpha$, for a method that has single-trial accuracy $p$ with computational effort $N$ is
\begin{eqnarray}
q_{\alpha}(p,N) &=& \nu(\alpha,p) N. 
\end{eqnarray}
We use this normalized computational effort $q(p,N)$ as a scoring function to compare different optimization methods, with $\alpha = 0.9$.
A low value of $q$ corresponds to a more efficient optimization algorithm.

The parameters of the cooling method (3 for classical SA; 8 for ACSA) fully determine the normalized effort $q$ for a particular system.
This is determined by performing $M$ independent trial optimizations to obtain $\left< N \right>$ and $\left< p \right>$, then using these to evaluate $q_{\alpha}\left( \left< p \right>, \left< N \right> \right)$.

In order to make a fair comparison between the new ACSA algorithm and the classical SA, the parameters of both methods were optimized to ensure they were as efficient as possible.
The Nelder-Mead downhill simplex method \cite{nel65} was used to optimize the parameters, using $M= 900$ trials for each point in parameter space to evaluate $\left< p \right>$ and $\left< N \right>$.




Once fully optimized parameters have been determined for both the classical SA and ACSA method on the same system, the efficiency of the ACSA method is defined using the ratio of their normalized computational efforts:
\begin{eqnarray}
\epsilon &=& \frac{q^{\mathrm{(SA)}}}{q^{\mathrm{(ACSA)}}}.
\end{eqnarray}
When ACSA can achieve the target accuracy with less computational effort than classical SA, the efficiency is greater than 1.

\begin{figure}
\includegraphics[width=\linewidth]{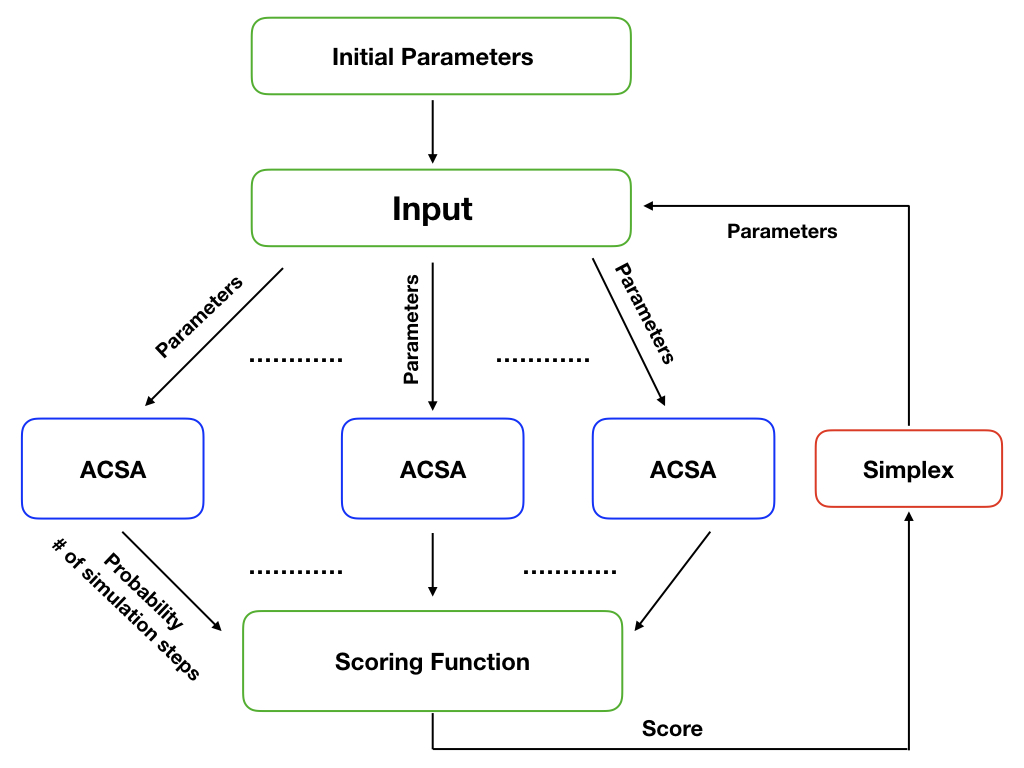}
\caption{\label{fig:methods} Parallel communication between simplex, adaptive-cooling simulated annealing, and the scoring function using MPI.}
\end{figure}

Accurate statistical estimation of $q$ requires multiple, independent optimizations to be performed.
Consequently, our implementation runs multiple simulated annealing optimizations in parallel. 
These parallel simulations are controlled, and their results are combined, using the message passing interface (MPI)\cite{gro94} communications protocol, allowing high-throughput calculations even with large $M$.
This communication is illustrated schematically in Fig.~\ref{fig:methods}%

The large-scale parallel replication of the optimizations is only needed here to perform a detailed analysis of the optimization method.
Production-level application of the ACSA method would not require $M=900$ repeated optimizations.
However, a smaller number of repeated optimizations may be called for, in cases where the optimal set of parameters is one that generates a quick-and-dirty optimization with relatively low $p$, and counts on boosting achieving additional accuracy through repeated trials.

The parameter optimization starts by running $M$ ACSA optimizations at each vertex of an initial simplex in parameter space. After all ACSA simulations have finished, the mean success probability and the total computational cost at each vertex is used to evaluate $q$. This score is sent back to the simplex code, which uses it as the objective function value at that vertex. Once values are available at each vertex, the simplex algorithm has enough information to generate a new vertex, which requires a new ACSA optimization at the new vertex. The simplex algorithm continues with simplex moves until the range of objective function values across the simplex is lower than a tolerance value of 0.001. The vertex with the lowest $q$ values is then taken to provide the optimal ACSA parameters.
This method could be used for optimizing the classical SA parameters, as well.
In practice, however, because the SA algorithm only has 3 parameters, that optimization was performed by a hand-directed grid search.

\section{Results}

To test the efficiency of the ACSA method, energy minimizations were performed on a number of Lennard-Jones (LJ) clusters, LJ$_n$.
This system has been well studied as a benchmark for optimization methods, and the minimum-energy structures are known\cite{wal97} for clusters with $n$ up to at least 1000.
Results will be discussed in detail for clusters with $n = 6,7,9,10,13,19,20,23,36$.
These clusters were chosen to span a range of cluster sizes and complexities, while still being small enough to explore with statistical detail.

Initial geometries of the LJ$_n$ clusters were obtained by placing the atoms randomly, with uniform distribution, within a distance of 2.74$\sigma$ of the origin.
The sampling steps of the simulated annealing algorithm were performed using molecular dynamics with the velocity Verlet thermostat\cite{Leach01}, using a timestep of 0.002 in reduced LJ units.
Temperature control during the equilibration and production phases was achieved using the Langevin thermostat\cite{Leach01}, with a friction coefficient of 0.002, also in reduced LJ units.

Because the heat capacities of LJ$_n$ clusters are well understood as a function of temperature,\cite{par10,fra95} the initial temperature for both the SA and ACSA optimizations was chosen to be a value that was somewhat above the peak in heat capacity.
All other parameters (7 for ACSA, 2 for SA) were optimized to minimize the value of $q_{0.9}$.
The optimized parameters for both methods are shown in Tables~\ref{tab:finalSA} and \ref{tab:finalACSA}.
The performance of the two methods with these optimized parameters is compared in Table~\ref{tab:performance}.

\begin{table}
\caption{\label{tab:finalSA}Optimized parameters for classical SA applied to energy minimization of LJ$_n$ clusters.}
\begin{ruledtabular}
\begin{tabular}{rddd}
$n$ & k & T_i\footnote{Initial temperature was set, not optimized.} & T_f \\
\hline
6& 4.50 \times 10^{-6} & 0.150 & 0.0020 \\
7& 5.39 \times 10^{-7} & 0.250 & 0.0354 \\
9& 6.50 \times 10^{-7} & 0.220 & 0.0395 \\
10& 3.73 \times 10^{-7} & 0.260 & 0.0869 \\
13& 6.16 \times 10^{-7} & 0.310 & 0.0867 \\
19& 3.52 \times 10^{-7} & 0.315 & 0.1299 \\
20& 7.28 \times 10^{-7} & 0.380 & 0.1087 \\
23& 9.38 \times 10^{-7} & 0.400 & 0.1385 \\
36& 1.15 \times 10^{-6} & 0.190 & 0.1180 \\
\end{tabular}
\end{ruledtabular}
\end{table}

\begin{table*}
\caption{\label{tab:finalACSA}Optimized parameters for ACSA applied to energy minimization of LJ$_n$ clusters.}
\begin{ruledtabular}
\begin{tabular}{rdddddddd}
$n$ & k_s & k_f & C_V^* & T_i\footnote{Initial temperature was set, not optimized.} &T_f & N_{\mathrm{eq}} & N_{\mathrm{prod}} & N_{\mathrm{cool}} \\
\hline
6 & 4.74\times10^{-6} & 1.35\cdot10^{-5} & 8.61 & 0.150 & 0.0015 & 91 & 75 & 96 \\
7 & 4.21\times10^{-7} & 1.06\times10^{-5} & 3.74 & 0.250 & 0.0175 & 79 & 81 & 70 \\
9 & 1.30\times10^{-6} & 8.81\times10^{-6} & 3.80 & 0.220 & 0.0330 & 284 & 332 & 102 \\
10 & 4.80\times10^{-7} & 1.91\times10^{-5} & 4.47 & 0.260 & 0.0326 & 78 & 94 & 70 \\
13 & 6.88\times10^{-7} & 7.33\times10^{-6} & 3.33 & 0.310 & 0.0617 & 95 & 250 & 103 \\
19 & 5.70\times10^{-7} & 1.85\times10^{-5} & 4.90 & 0.315 & 0.0723 & 30 & 47 & 117 \\
20 & 8.64\times10^{-7} & 9.12\times10^{-6} & 5.83 & 0.380 & 0.0656 & 55 & 37 & 92 \\
23 & 1.06\times10^{-6} & 1.39\times10^{-4} & 5.96 & 0.400 & 0.0466 & 59 & 62 & 125 \\
36 & 1.04\times10^{-6} & 6.30\times10^{-5} & 4.35 & 0.190 & 0.1473 & 35 & 28 & 70 \\
\end{tabular}
\end{ruledtabular}
\end{table*}

\begin{table*}
\caption{\label{tab:performance}Performance of the classical SA and ACSA algorithms for energy minimization of  LJ$_n$ clusters\cite{For17,sch97}.}
\begin{ruledtabular}
\begin{tabular}{rrddddddd}
 & & \multicolumn{3}{c}{SA} & \multicolumn{3}{c}{ACSA} & \\
 \cline{3-5}
 \cline{6-8}
$n$ & $N_{\mathrm{min}}$ & \left< p \right> & \left< N \right> & q & \left< p \right> & \left< N \right> & q & \epsilon \\
\hline
6 & 2 & 0.568 
& 9.59\times10^5 
& 9.59\times10^{6} & 0.540 
& 8.61\times10^5 
& 1.00\times10^{7} & 0.95\\
7 & 4 & 0.9056
& 3.63\times10^6 
& 1.08\times10^{7} & 0.799 
& 1.59\times10^6 
& 9.34\times10^{6} & 1.16\\
9 & 21 & 0.9556 
& 2.64\times10^6 
& 5.28\times10^{6} & 0.9567 & 1.49\times10^6 
& 4.37\times10^{6} & 1.21\\
10 & 64 & 0.9144 
& 2.94\times10^6 
& 8.82\times10^{6} & 0.693 
& 1.22\times10^6 
& 9.03\times10^{6} & 0.98\\
13 & 1510 & 0.9067 
& 2.07\times10^6 
& 6.20\times10^{6} & 0.9100 & 1.35\times10^6 
& 5.00\times10^{6} & 1.24\\
19 & $\sim2\times10^{6}$ & 0.706 
& 2.51\times10^6 
& 1.50\times10^{7} & 0.698 
& 1.29\times10^6 
& 9.53\times10^{6} & 1.55\\
20 & & 0.712 
& 1.72\times10^6 
& 1.03\times10^{7} & 0.793 
& 1.50\times10^6 
& 8.79\times10^{6} & 1.17\\
23 & & 0.268 
& 1.13\times10^6 
& 2.60\times10^{7} & 0.550 
& 1.03\times10^6 
& 1.21\times10^{7} & 2.15\\
36 & & 0.158 
& 5.80\times10^5 
& 1.94\times10^{7} & 0.257 
& 2.46\times10^5 
& 7.29\times10^{6} & 2.67\\
\end{tabular}
\end{ruledtabular}
\end{table*}

\begin{figure}
\includegraphics[width=\linewidth]{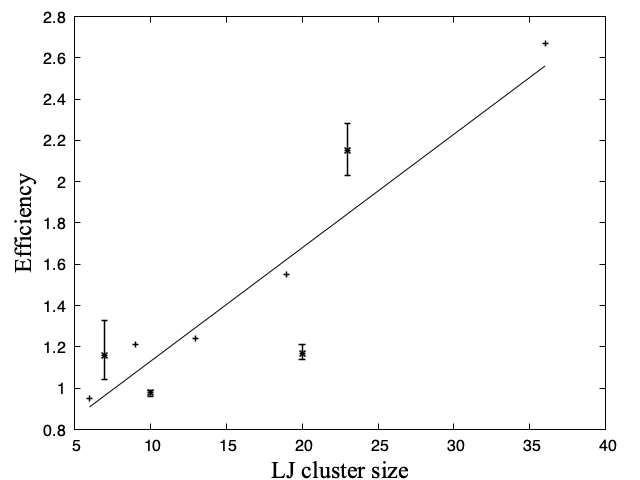}
\caption{\label{fig:eff} Efficiency of the ACSA over the classical SA for LJ clusters.
Error bars represent the standard error of the mean.
The straight line is a least squares fit, but is only intended as a guide to the eye.}
\end{figure}

Fig.~\ref{fig:eff} shows the efficiency of the ACSA algorithm, relative to classical SA, and how this efficiency depends on cluster size.
Although the methods are comparable in efficiency at small cluster size, the ACSA algorithm begins to perform better as the cluster size increases.
In particular, the ACSA algorithm is able to find the minimum-energy cluster more than twice as fast as classical SA (even after accounting for the overhead in evaluating the heat capacity), once the cluster size exceeds $n=20$.
It is reasonable to expect that this advantage will continue to increase for larger, more computationally demanding optimizations.

\begin{figure}
\includegraphics[width=\linewidth]{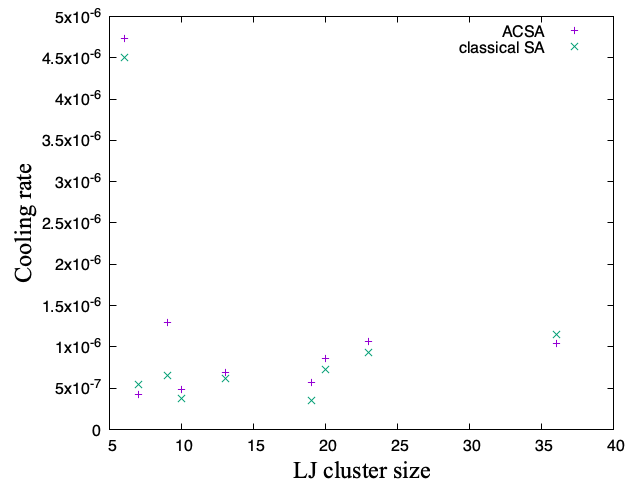}
\caption{\label{fig:Ks} Optimal classical SA cooling rates ($k$) and ACSA slow cooling rates ($k_{\mathrm{s}}$) for a number of LJ cluster sizes.}
\end{figure}

\begin{figure}
\includegraphics[width=\linewidth]{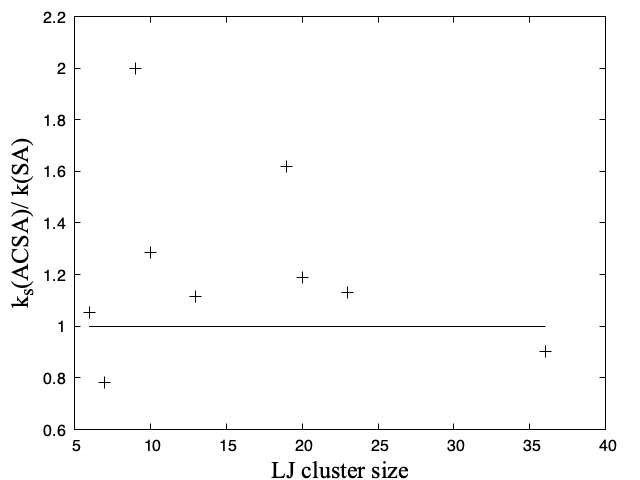}
\caption{\label{fig:sACvsSA} Slow cooling rates for ACSA and classical SA for LJ clusters.}
\end{figure}

It is interesting to explore the reasons for the computational advantage of the ACSA algorithm.
Fig.~\ref{fig:Ks} compares the ACSA slow cooling rate, $k_{\mathrm{s}}$, to the optimal SA cooling rate at each cluster size.
These cooling rates are quite similar, thus confirming the initial motivation for the method:
The classical SA algorithm performs most efficiently with a cooling rate that provides a balance between low error rates and fast optimization; the ACSA algorithm independently uses very nearly the same optimal cooling rate, but only in the crucial region of phase space where the number of thermally accessibly states is decreasing rapidly.
This point is emphasized further in Fig.~\ref{fig:sACvsSA}, which shows that the ratio $k_{\mathrm{s}}(\mathrm{SA}) / k(\mathrm{ACSA})$ is always relatively close to 1, even though the magnitude of the individual $k$ values varies by as much as a factor of 10 for different clusters.

\begin{figure}
\includegraphics[width=\linewidth]{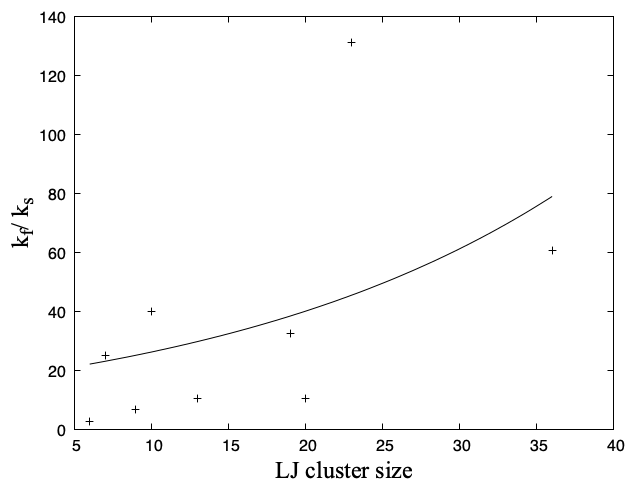}
\caption{\label{fig:fs} Ratio of the fast cooling rate to the slow cooling in ACSA for several cluster sizes.}
\end{figure}

The classical SA method cools at the same rate as it anneals through all regions of phase space, even those regions that have relatively low risk of quenching into local minima.
The ACSA algorithm, on the other hand, has the flexibility to cool at a faster rate when this kinetic trapping risk is low.
Fig.~\ref{fig:fs} shows the ratio $k_{\mathrm{f}} / k_{\mathrm{s}}$ for the ACSA algorithm.
The fast cooling rates are always at least several-fold faster than the slow cooling rates, even for the smallest clusters.
This ratio increases with increasing cluster sizes, exceeding a factor of 50 for the larger clusters ($n>20$)  where ACSA is most efficient.

Thus, the speedup obtained by the ACSA algorithm results from these periods of faster cooling.
The cooling is nearly the same as classical SA in the crucial bottleneck regions of the energy landscape, but much faster at other times.
The computational advantage of this faster cooling is more than enough to make up for any increased error rate, as well as well as the computational overhead associated with evaluating the heat capacity.


This point is reinforced by examining how the values of $\left< p \right>$ and $\left< N \right>$ differ for ACSA from the values for classical SA (see Table~\ref{tab:performance}).
The optimal ACSA value of $\left< p \right>$ is in some cases larger than the value for SA, and in some cases smaller. 
But the ACSA algorithm always succeeds in finding the minimum more quickly, with a smaller value of $\left< N \right>$.
Regardless of whether the modified cooling rates result in a more or less accurate optimization, the benefit comes from reducing the effort required to reach the answer.

\begin{figure}
\includegraphics[width=\linewidth]{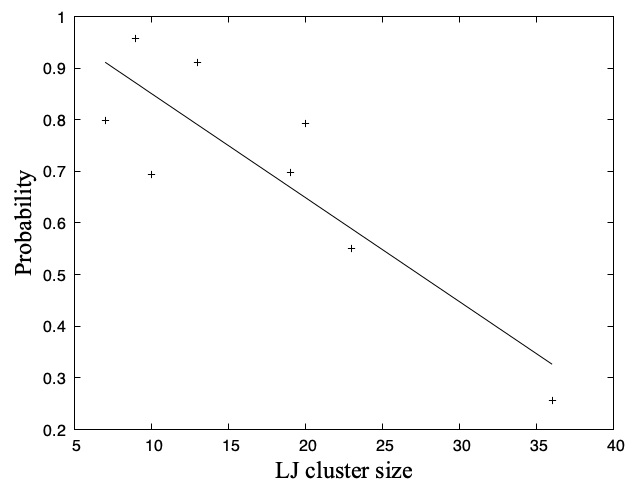}
\caption{\label{fig:Prob} Probability of successfully reaching the global minimum as a function of LJ cluster size.}
\end{figure}

The (optimal) probability of minimizing into the correct global minimum decreases as the cluster size increases, as can be seen from both Table~\ref{tab:performance} and Fig.~\ref{fig:Prob}.
This is not surprising, as it is due to the rapid increase in the number of local minima with increasing dimensionality of the energy landscape.
(Table~\ref{tab:performance} lists the number of local minima\cite{For17,sch97} for the smaller clusters.
This value rises steeply enough that the energy landscape has not been fully explored for even moderately large clusters.)
With the exception of the $n=6$ cluster (which has only two local minima, and can tolerate an anomalously fast cooling rate and correspondingly poor success rate), most of the clusters have an increasingly hard time finding the global minimum as the cluster size increases.

Table~\ref{tab:finalACSA} also lists the optimized number of equilibration, production, and cooling steps chosen for the ACSA algorithm.
There is considerable variation in these values, but as a general observation it appears that the number of steps in each block is roughly equal.
Averaging over all cluster sizes, 30\% of the steps are spent equilibrating, 38\% in production evaluating the heat capacity, and 32\% cooling the system.
Thus, roughly a third of the computational effort is spent actually annealing the system, with about two thirds of the computational effort spent on the overhead required to evaluate the heat capacity.

One advantage of the classical SA algorithm over the new ACSA algorithm is its smaller number of parameters (1 cooling rate and 3 total parameters, vs 2 cooling rates and 8 total parameters).
The data presented so far has been for uses of the algorithms (both SA and ACSA) with a set of parameters that has been fully optimized for each cluster size.
The computational effort required to optimize these parameters, of course, is several orders of magnitude larger than the computational effort required for a single optimization.
Thus, the computational advantage of the ACSA algorithm over classical SA will only be useful if it can be implemented without requiring a full parameter optimization.
Thus, we also performed several tests to gauge the success of the ACSA algorithm on LJ clusters without the benefit of a full parameter optimization.
In the first of these tests, the ACSA parameters were chosen based only on the  values of the (optimal) classical SA parameters.
In this test, the classical SA algorithm has the advantage of using optimized parameters, but the ACSA parameters are chosen using heuristics so as not to involve any additional computational effort.
The previously studied LJ$_{23}$ cluster was used for this test.
In the second test, both the SA and ACSA parameters were chosen using heuristics, for a previously unstudied cluster.
The LJ$_{24}$ cluster was used in this case.

Several rules of thumb for determining ACSA parameters were suggested by the preceding results.
Fig.~\ref{fig:sACvsSA} suggests the heuristic that
\begin{eqnarray}
k_s &=& \alpha k,
\end{eqnarray}
with $\alpha = 1.23$.
That is, the ACSA slow cooling rate is taken to be nearly the same as the SA cooling rate (slightly larger, based on the cases examined so far).


Similarly, 
\begin{eqnarray}
k_f &=& n^\beta k,
\end{eqnarray}
where $n$ is the LJ cluster size and we take $\beta = 1.139$. 
That is, the ACSA fast cooling rate increases with cluster size, as seen in Fig.~\ref{fig:fs}.
The initial temperature is taken to be the same for both SA and ACSA ($T_i = 0.4$), for LJ$_{23}$.
As with all clusters, this was taken to be a temperature somewhat above the observed peak in heat capacity.\cite{par10,fra95}
The final temperature for ACSA is taken to be half that of the final temperature for the classical SA, as a relatively conservative example of the behavior seen in Tables~\ref{tab:param1} and \ref{tab:finalSA}), where the optimal $T_{\mathrm{f}}$ for ACSA is lower than the optimal $T_{\mathrm{f}}$ for classical SA.
Presumably this occurs because the ACSA algorithm is cooling at the fast rate in the late stages of the optimization, and thus does so with less computational cost, pushing the balance towards the slightly higher accuracy achieved by cooling more thoroughly.
The heat capacity cutoff, $C_V^*$, is chosen by observing that the optimal value of $C_V^*$ for the other clusters is roughly half of the peak heat capacity value for that LJ cluster.\cite{par10,fra95}
For the LJ$_{23}$ cluster, this corresponds to a value of $C_V^* = 6.0$, in reduced units. 
Lastly, the number of steps in the equilibration, production, and cooling phases were chosen using
\begin{eqnarray}
N_{Cool} &=& \gamma \cdot (N_{Eq} + N_{Prod})
\end{eqnarray}

For the LJ\textsubscript{23} cluster optimization we chose $\gamma$ of 0.73 to predict the number of cooling optimization steps, which is a mean value  $\gamma$ for previously optimized clusters. On average, the simulations require about 1.2 times more production steps than equilibration simulation steps, and the ratio slowly decreases for the clusters larger than LJ\textsubscript{13}. The number of equilibration steps for LJ\textsubscript{23} is a mean value of the equilibration steps for previously optimized clusters.


\begin{table*}
\caption{\label{tab:param1}Predicted (non-optimized) ACSA parameters for the LJ$_{23}$ and LJ$_{24}$ cluster optimizations.}
\begin{ruledtabular}
\begin{tabular}{rdddddddd}
$n$ & k_s & k_f & C_V^* & T_i &T_f & N_{\mathrm{eq}} & N_{\mathrm{prod}} & N_{\mathrm{cool}} \\
\hline
23 & 1.15\times10^{-6}&3.33\times10^{-5}&6.0&0.4&0.069& 90& 108& 145\\
24 & 1.02\times10^{-6}&6.5\times10^{-5}&5.6&0.4&0.12&45&50&114\\
\end{tabular}
\end{ruledtabular}
\end{table*}

The parameters predicted by these heuristics are summarized in Table~\ref{tab:param1}, and the performance of the resulting optimization is summarized in Table~\ref{tab:pred_perf}.
Even without optimized parameters, the ACSA algorithm is 1.06 times more efficient than the (fully optimized) classical SA algorithm.
This is not as good as the efficiency of $\epsilon = 2.15$ value achieved with optimized parameters, but it illustrates that the ACSA algorithm can outperform SA without any additional effort spent on tuning the ACSA performance.  

\begin{table*}
\caption{\label{tab:pred_perf}Performance of the classical SA and ACSA algorithms for two test cases with heuristically predicted parameters.}
\begin{ruledtabular}
\begin{tabular}{rddddddd}
 & \multicolumn{3}{c}{SA} & \multicolumn{3}{c}{ACSA} & \\
 \cline{2-4}
 \cline{5-7}
$n$ & \left< p \right> & \left< N \right> & q & \left< p \right> & \left< N \right> & q & \epsilon \\
\hline
23 & 0.268 & 1.13\times10^6 & 2.60\times10^7 & 0.438 & 1.75\times10^6 & 2.45\times10^7 & 1.06 \\
24 & 0.373 & 1.30\times10^6 & 2.98\times10^7 & 0.361 & 1.18\times10^6 & 2.72\times10^7 & 1.10
\end{tabular}
\end{ruledtabular}
\end{table*}

It is more typically the case that the optimal SA parameters are not known either, and are estimated heuristically, based on trial and error, or past experience.
For the LJ$_{24}$ cluster, where the optimal parameters are not known, we estimated the parameters for both SA and ACSA, based on the previous optimizations.

The cooling rate for classical SA was chosen to be $k = 8.9 \times 10^{-7}$, based on the trend line in Fig.~\ref{fig:Ks}.
The initial and final temperatures were chosen to be $T_{\mathrm{i}} = 0.38$ and $T_{\mathrm{f}} = 0.12$, positioning them to either side of the temperatures at which the heat capacity is observed to have a maximum.\cite{par10,fra95}.

For ACSA, the estimated SA parameters were combined with the previously developed heuristics to obtain estimated ACSA parameters.
The ACSA parameters are summarized in Table~\ref{tab:param1}, and the performance of both SA and ACSA algorithms is summarized in Table~\ref{tab:pred_perf}.
This test case is more representative of a practical optimization, where the optimization parameters are not fully optimal, but are only roughly refined through experience with previous optimizations.
Neither algorithm performs as well as for the LJ$_{23}$ case, with optimized SA parameters.
But the ACSA still has an advantage over SA, completing the optimization 1.1 times more efficiently.

\section{Conclusions}

We have introduced a new global optimization method, which we call adaptive-cooling simulated annealing (ACSA), in which the SA cooling rate varies as the optimization proceeds, depending on the current heat capacity of the system.
By using comparable cooling rates to traditional SA only when the system is annealing through the bottleneck region of phase space, and faster cooling rates at other times, the optimization proceeds more efficiently --- more than compensating for the extra computational cost of evaluating the heat capacity.

The method has been demonstrated for small LJ clusters.
It is comparable in efficiency to traditional SA for the smallest clusters, becoming more than twice as efficient for LJ$_n$ clusters with $n>20$, and with an efficiency that rises as the cluster size increases.
It is reasonable to expect that the same trend will also apply to other systems as well: more complex energy landscapes will benefit more by automatically detecting the regions in which cooling can be done more rapidly.

The ACSA algorithm requires the choice of several additional parameters over classical SA.
For the specific case of LJ clusters, a set of heuristics were determined for obtaining reasonable optimization parameters, which have proven to result in improvements over SA.
In the general case, ACSA can exactly reproduce the SA algorithm by using $k_s = k_f = k$ and $N_{\mathrm{eq}} = N_{\mathrm{prod}} = 0$.
Thus, the SA performance is available as a lower bound.
Then $C_V^*$ can be chosen based on values observed during a trial observation, and $k_f$ can be increased in order to gain efficiency, investing as much or as little effort into optimizing the parameters as is justified for the particular application.

In this proof-of-concept demonstration of the method, the algorithm was kept rather simple.
It is easy to imagine extensions which will make the method even more efficient, however.

First of all, the heat capacities can be calculated on-the-fly during the cooling phase without introducing the prior equilibration and production phases. This will reduce the computational cost by approximately a factor of three.
Another possible extension is an adaptive cooling schedule that uses a more general $K(T)$ function, rather than a simple switch between two discrete values. 
Lastly, it will likely be beneficial to use the density of states, rather than the heat capacity, to determine the cooling rate.
The heat capacity provides a useful proxy for the number of states accessible at a particular temperature, but it would be more valuable to detect the number of {\em barriers} accessible.
This is more closely related to the density of states.
This property is more difficult to calculate, but could prove a more efficient statistical mechanical indicator for the optimal cooling rate.
These extensions to the ACSA method will be explored in future investigations. 

\nocite{*}
\bibliography{aipsamp}

\end{document}